# Temperature Control of Spin-Wave Spectra in Continuously Graded Epitaxial Pd-Fe Alloy Films

I.V. Yanilkin[a*], A.I. Gumarov[a,b], B.F. Gabbasov[a], R.V. Yusupov[a], L.R. Tagirov[b,c]

[a] *Institute of Physics, Kazan Federal University, Kazan, 420008 Russia*
[b] *E.K. Zavoisky Physical-Technical Institute of RAS, Kazan 420029, Russia*
[c] *Moscow Institute of Physics and Technology, Dolgoprudny, Moscow 141700, Russia*

[*]Correspondence e-mail: yanilkin-igor@yandex.ru

**Abstract**

Continuously graded ferromagnetic thin films represent a new class of magnonic materials, where a spin-wave resonance spectrum can be tuned in resonance frequencies/fields, number of modes and even dispersion law by the magnetic properties profiling across the film thickness. In the paper, we demonstrate that temperature is another degree of freedom controlling the spin-wave spectra in graded magnetic materials. We show that temperature affects the spectrum not only in a trivial way via the temperature dependence of magnetization and related quantities, but rather through modification of magnetic structure of a film due to the Curie temperature being crossed by fractions of its thickness. The profile that is continuous at lower temperatures may become discontinuous at higher temperatures. To demonstrate this, four 200-nm thick vertically graded epitaxial $Pd_{1-x}Fe_x$ films were synthesized using molecular beam epitaxy technique, in particular, two with the linear (2-10 at.% and 12-18 at.% range of the iron content), one with the sine and one with the cosine (both with the 2-10 at.% range of the iron content) distribution profiles. The resonance spectra of standing spin waves were studied by the cavity ferromagnetic resonance at 9.4 GHz in the temperature range of 20-300 K. Versatile spin-wave excitation patterns were obtained, and drastic evolution of their excitation energies, and number of modes with temperature was observed. The standing spin-wave resonance spectra were modeled using a classic Landau-Lifshitz-Gilbert approach, which demonstrated excellent agreement with the experimental data for all samples in the entire temperature range. Based on the modeling, temperature dependences of the spin-wave stiffness $D$ and magnitudes of the interface $\alpha_{inter}$ and surface $\alpha_{surf}$ pinning constants were obtained. Our study shows that in addition to a continuous grading of magnetic properties of thin films, temperature provides another powerful and well elaborated tool for modifying the magnetic grading profile and thus, flexile tuning the spectrum of spin-wave excitations.

**Introduction**

Magnonics is a rapidly developing field of physics and technology covering studies and applications of spin waves for computing and information processing (see, for example, [1-8] and references therein). Graded magnetic thin films (films with spatially varying magnetic properties [9]) are promising materials for magnonics due to a nonreciprocity of propagating spin waves [10] and a possibility to design the spectrum of standing spin waves [11-16]. In the latter case, the magnetization profile determines both intensities and spectrum patterns of the standing spin wave resonances (SSWR). Recently, we succeeded in synthesis of high-quality epitaxial films of palladium-iron alloy $Pd_{1-x}Fe_x$ with the iron content $x$ varying in the range of 0.01-0.5 [17-19]. The alloy is a metallic ferromagnet, magnetic properties of which vary in a wide range depending on the iron content $x$ [18,20]. Spontaneous magnetization of $Pd_{1-x}Fe_x$ alloys appears already at very low concentration of iron above ~ 0.01-0.1 at.% with Curie temperature of several K and a "giant" effective magnetic moment per Fe atom [21]. Such diluted alloys with the iron content in the interval of 1-4% are of particular interest in superconducting spintronics [22-30]. For higher concentrations, the magnetization and the Curie temperature increase approximately linearly with increasing Fe concentration (see [18, 20] and text below), and the interval of 2-20 at.% is suitable for the use in the low-temperature to room-temperature magnonics, because magnetic properties of the alloy can be easily tuned for particular application.

With our original technology of programming the time course of deposition rates of Pd and Fe effusion cells [31], we extended the earlier concept of the Pd-Fe alloy dilution towards a synthesis of magnetically graded epitaxial films with sophisticated predetermined magnetization profiles across a film thickness [32,33]. The investigation of the SSWR spectra of these magnetically graded films performed at a low temperature of $T = 20$ K, indeed confirmed the ability to engineer an SSWR dispersion law and their energy/frequency range.

Besides the composition of graded films of Pd-Fe alloy, temperature significantly affects their spin-wave properties. Moreover, temperature variation influences SSWRs not only in a trivial way via a respective dependence of magnetization and related parameters [18,20]. Rather, in a case of a large amplitude of the iron concentration variation, the temperature rise causes a transition of low-Curie-temperature fractions of a film to paramagnetic state thus controllably switching a continuous at lower temperatures magnetization profile to a discontinuous one. This of course dramatically changes the structure of the spin-wave excitation pattern and its energy range. Temperature, therefore, serves as a complementary and, importantly, easily and accurately controlled degree of freedom for modification of the magnetic profile of a film for tuning the spectrum of spin-wave excitations. The goal of our work was to study the temperature evolution of the SSWR spectra in various continuously graded epitaxial Pd-Fe alloy films in a broad temperature range covering the cases of crossing the Curie temperature by a fraction of a film thickness. For the study, four magnetically graded samples were fabricated: two with a linear, one with a sine, and one with a cosine magnetization profile. The results show, first, that moderate variations of temperature significantly alter the SSWR spectrum in terms of a number of excited modes and their energy. Second, the Landau-Lifshitz-Gilbert model successfully describes the experimental results for all samples over the entire temperature range of spectra observation. Third, temperature dependences of the spin stiffness coefficient and pinning strength at film boundaries were obtained.

**Samples and experimental details**

Graded epitaxial films of Pd-Fe alloy were produced by means of palladium and iron metals evaporation from effusion cells in the form of molecular beams and their deposition onto a rotating single-crystal MgO (001) substrate. To obtain magnetically inhomogeneous across the thickness films, the deposition rate of palladium was kept constant while that of iron was varied in a controlled manner. In total, four graded epitaxial films of a palladium-iron alloy were obtained, in particular, two with a linear, one with a sine and one with a cosine iron distribution profile. Each synthesized sample was about 200 nm thick. Samples with linear profiles differed in concentration ranges: one covered the range of 2-10 at.% (hereafter referred to as Lin_2-10), the other – 12-18 at.% (Lin_12-18). In samples with a sine (Sin_2-10) and a cosine (Cos_2-10) profiles, the iron concentration range was 2-10 at.% with the variation period equal to the film thickness. In addition, homogeneous epitaxial Pd-Fe films with iron concentrations of 1, 2, 4, 7, 8, and 12 at.% and a thicknesses of 100-400 nm were fabricated for pilot studies and verification of a model. Details of the synthesis method and structural studies are described in Refs. [17-19,31].

Spin wave resonances in the films were studied with the Bruker ESP300 continuous-wave X-band spectrometer in the temperature range of 10 – 300 K. The temperature dependence of the saturation magnetization and the magnetic hysteresis loops were measured by the vibrating sample magnetometry with the Quantum Design PPMS-9 system. Film thicknesses were measured with the Bruker Dektak XT stylus profilometer by the shadow mask method.

**Theoretical background**

Standing spin wave resonances in ferromagnetic film with the magnetization profile $M_s(z,T)$ across the thickness $z$ are described by the equation for circular projection of magnetization $m(z,T)$ [11,32-34]:

$$\left[-D(T)\frac{\partial^2}{\partial z^2} + V(z,T)\right]m_n(z,T) = -H_n^{res}m_n(z,T). \qquad (1)$$

Here, $D(T) = \frac{2A(T)}{\mu_0 M_s(T)}$ is the spin stiffness magnitude, simplified to be independent on $z$ and determined only by temperature $T$, and $V(z,T)$ is a crucial parameter, analogous to a "potential well" in the Schrödinger equation and determining the filling of the film thickness with standing spin waves

$$V(z,T) = -\frac{2\pi f_{res}}{\gamma} - \beta(T)M_s(z,T) + \frac{D(T)}{M_s(z,T)}\frac{\partial^2 M_s(z,T)}{\partial z^2}. \qquad (2)$$

Here $f_{res}$ is the excitation frequency, $\gamma$ is the gyromagnetic ratio, $\beta(T) = \frac{M_{eff}(T)}{M_s(T)}$ is the ratio of the effective magnetization to the saturation magnetization, also simplified to be independent on the $z$ coordinate. The surface pinning is determined by the boundary conditions

$$\frac{dm}{dz} + \alpha_s m = 0, \qquad (3)$$

where the surface pinning coefficient $\alpha_s = K_s/A_s$ is the ratio of the surface energy to the surface exchange stiffness. Parameter $\alpha_s$ was determined for each surface/interface of the ferromagnetic part of the film. Integral intensities of Lorentzian-shape resonance lines were found from modelling while their widths were taken equal to those observed in the experiment.

**Experiment and Modeling Results**

In Figure 1, the SSWR spectrum of the Lin_2-10 sample measured at $T$ = 10 K is presented. The experimental spectrum for each temperature $T$ was fitted by adjusting four parameters: $D(T)$, $\beta(T)$, $\alpha_{\text{surf}}(T)$ and $\alpha_{\text{inter}}(T)$, where the last two are the pinning coefficients for the film surface and the interface with the substrate, respectively. The four panels of Fig. 1 demonstrate how the SSWR spectrum pattern responds to variations in the above parameters. As one can see in Fig. 1a, the spin stiffness parameter $D$ determines the field gap between adjacent resonances: the larger the $D$ value, the wider the gap. Parameter $\beta$ variation (Fig. 1b) leads mainly to a shift of the entire SSWR spectrum as a whole: depending on whether it is greater or less than unity, the spectrum shifts towards higher or lower fields, respectively. Variations in the pinning parameters have a sufficiently weaker effect (Fig. 1c). A real iron concentration profile in a film may depart from a predefined one within 0.5 at.% [31]. The latter significantly affects the spin wave spectrum, as shown in Figure 1d. All these factors were taken into account when estimating the uncertainties in temperature dependences of $D(T)$, $\beta(T)$, $\alpha_{\text{surf}}(T)$ and $\alpha_{\text{inter}}(T)$. Other important parameters, such as the film thickness and the shape of concentration profile, were not adjusted in the course of the spectra modeling.

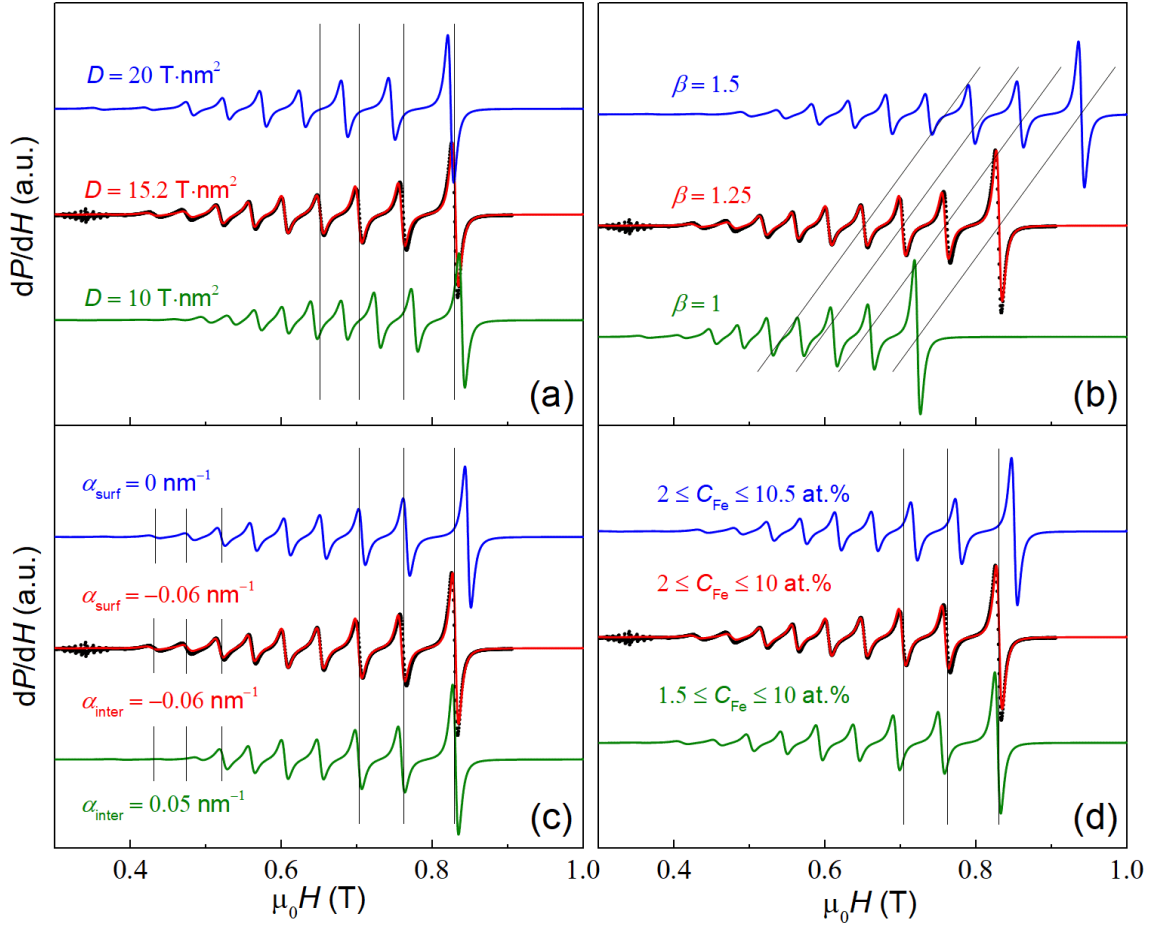

**Figure 1.** Spectrum of standing spin waves at $T = 10$ K of the Lin_2-10 sample (black dots), its fit (red line) and its modifications (blue and green lines) with the model parameter variation: (a) – spin stiffness $D$, (b) – ratio of effective magnetization to saturation magnetization $\beta$, (c) – pinning coefficients $\alpha_{\text{surf}}$ and $\alpha_{\text{inter}}$, (d) – iron concentration range of the linear profile.

In the Pd-Fe alloy, the paramagnetic Pd-matrix becomes spin-polarized with a magnetic moment of about 0.15-0.3 $\mu_B$/atom [35] due to a hybridization with iron atoms. With increasing the iron concentration, the saturation magnetization $M_s$ of the alloy and the Curie temperature $T_C$ increase (Fig. 2). In the work of Ododo [36], based on the analysis of a large volume of experimental data for bulk Pd-Fe samples, the following empirical expressions for $M_s$ and $T_C$ concentration dependences were obtained:

$$M_s(c_{\text{Fe}}) = 0.104 \cdot (c_{\text{Fe}}(\text{at.\%}) - 0.112)^{0.75} \, [\mu_B], \tag{4}$$

$$T_C(c_{\text{Fe}}) = 44.44 \cdot (c_{\text{Fe}}(\text{at.\%}) - 0.112)^{0.725} \, [\text{K}], \tag{5}$$

for $0.8 \leq c_{\text{Fe}} \leq 16$ at.%. The properties of homogeneous epitaxial Pd-Fe films synthesized by us follow well these dependences (Fig. 2b). Moreover, the characteristic dependence of the reduced magnetization $M_s/M_{s0}$ on the reduced temperature $T/T_C$ of the obtained epitaxial films is universal (Fig. 2a) for $1 \leq c_{\text{Fe}} \leq 12$ at.% and is described by the Kuz'min formula [37] with parameters $k = 1.8$ (dimensionless ratio of the fourth order to the second order coefficients in the Landau expansion for the free energy in powers of $M_s$) and $p = -0.02$ (provides a correct low-temperature behavior of the magnetic moment $M_s(T)$). The disclosed universal character of $M_s(T/T_C)/M_{s0}$ and Eqs. (4) and (5) were used in the modeling of magnetization profiles of the graded films at various temperatures for solving Eq. (1).

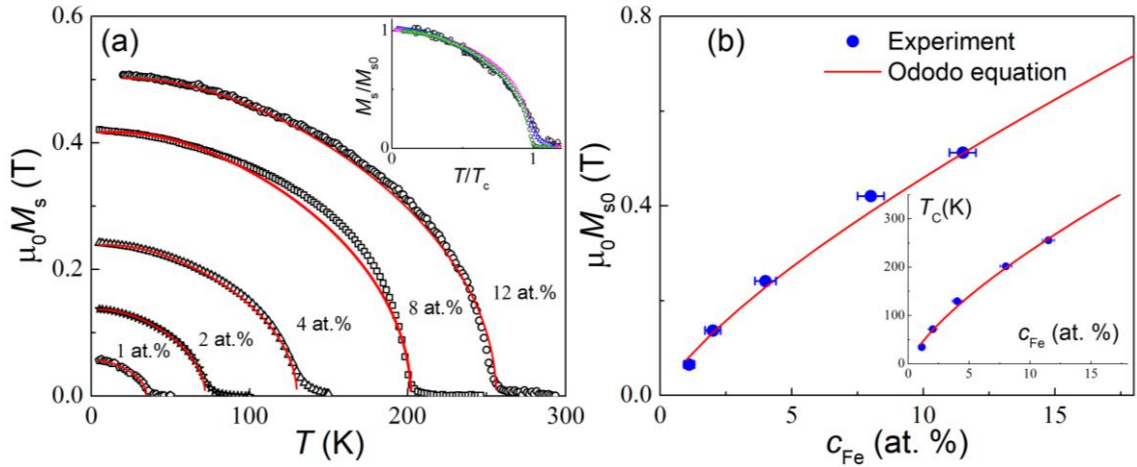

**Figure 2.** (a) Temperature dependences of the saturation magnetization of homogeneous Pd-Fe alloy films with different iron concentrations. The lines were obtained using the Kuz'min formula with the parameters $k$ and $p$ given in the text body. The inset shows the same dependences in reduced values, demonstrating the approximate universality of the temperature dependences of saturation magnetization; (b) Dependences of the saturation magnetization at 5 K and the Curie temperature on the iron concentration for homogeneous Pd-Fe films.

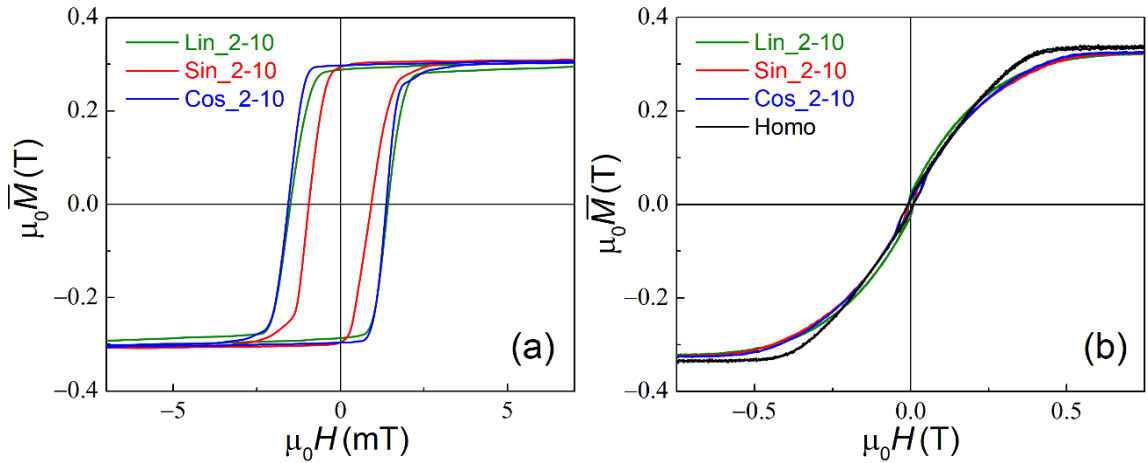

**Figure 3.** Magnetic hysteresis loops for the samples Lin_2-10, Sin_2-10 and Cos_2-10 measured in an external magnetic field parallel to the film plane (a) and along the normal to the film surface (b). The sample labeled "Homo" is a homogeneous epitaxial Pd-Fe alloy film with an iron concentration of 7 at. % ($Pd_{0.93}Fe_{0.07}$). All loops were recorded at 20 K.

All synthesized graded samples are easy-plane magnetic systems with small coercive fields (Fig. 3). Significant nonlinearity of the out-of-plane magnetic hysteresis curves indicates a noncollinear rotation of magnetic moments at different depths of the films (Fig. 3b). Regions of low magnetization line up along the field at lower field values compared to regions with higher magnetization. Now we switch to presentation and modeling of the temperature variations of SSWR in individual graded epitaxial Pd-Fe films.

*Sample Lin_2-10*

Figures 4 a,c show the temperature dependences of the standing spin-wave spectrum of the Lin_2-10 sample. As one can see, with increasing temperature, the resonances shift towards lower fields, and the number of SSWR modes decreases. Such evolution of the SSWR spectrum is qualitatively explained by the fact that with increasing temperature, the local magnetization of the film decreases, and a region of Pd-Fe with low iron content becomes paramagnetic. Thus, the actual thickness of the ferromagnetic part of the film decreases (Fig. 4b).

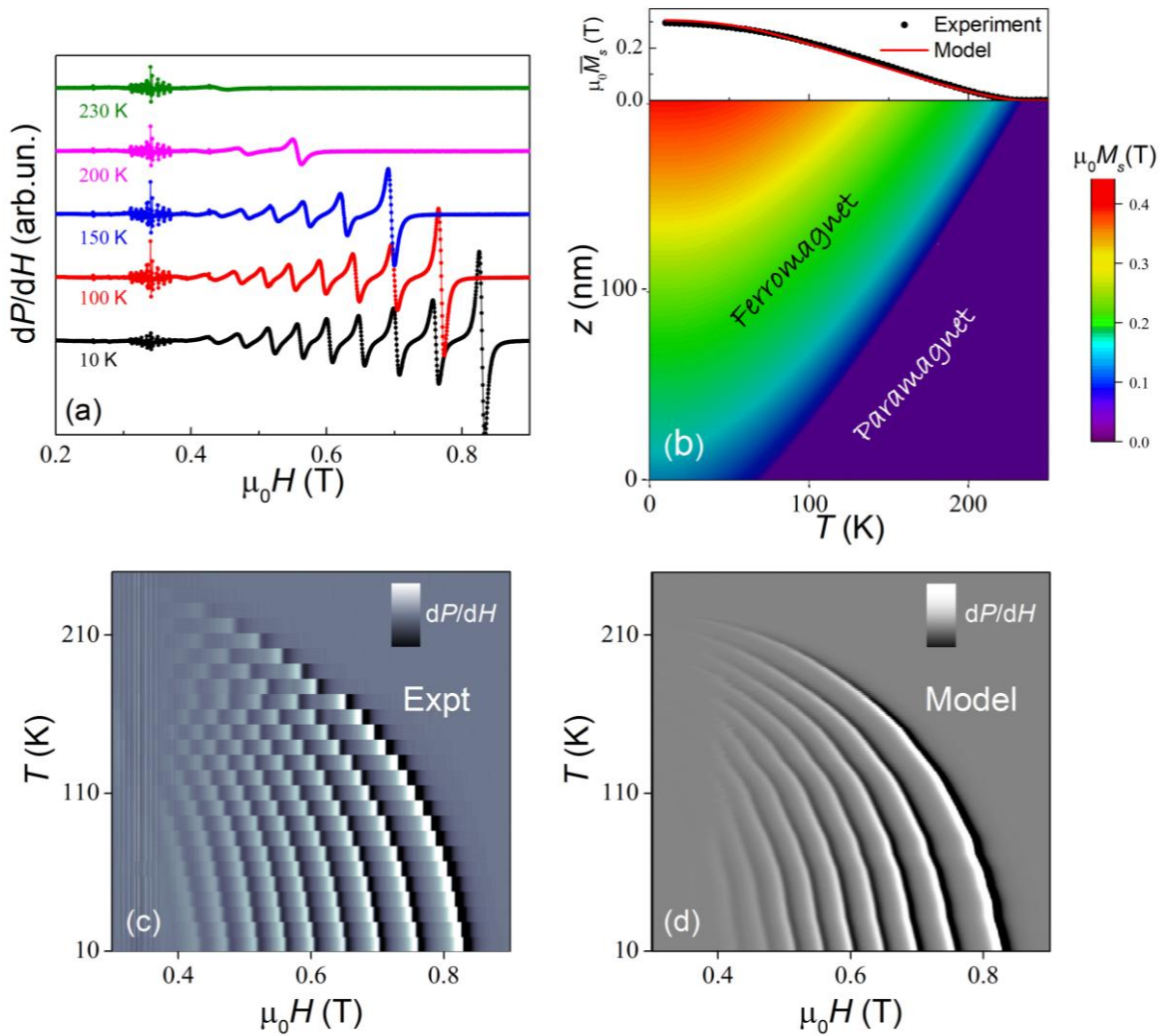

**Figure 4.** Variation of the experimental SSWR spectrum of the Lin_2-10 sample with temperature in the conventional representation (a) and as a contrast map (c). A set of sharp lines in the field region of ~ 0.34 T originates from the electron paramagnetic resonance of impurities in the MgO substrate. Distribution of saturation magnetization across the film thickness at various temperatures and the calculated temperature dependence of the mean magnetization in comparison with the experimentally measured one (b). Simulated temperature dependence of the spin-wave resonance spectrum (d).

Examples of the SSWR spectra modeling are shown in Figure 5a. Clearly, the calculated spectra reproduce well the experimental data (Figs. 5a, 4c and 4d). This demanded the parameter $\beta$ to be larger than 1 and weakly dependent on temperature (Fig. 5b). Note that for homogeneous epitaxial Pd-Fe films with iron concentrations in the range of 4-10 at.%, $\beta$ is larger than 1 based on our FMR studies (see Supplementary material). The surface pinning coefficients for the film surface $\alpha_{\text{surf}}(T)$ and the interface with substrate $\alpha_{\text{inter}}(T)$ have small values (weak pinning), however, they exhibit significantly different temperature dependences (Fig. 5c). Pinning at the surface $\alpha_{\text{surf}}(T)$ varies gradually over the entire temperature range. On the contrary, $\alpha_{\text{inter}}(T)$ at low temperatures is approximately equal to $\alpha_{\text{surf}}(T)$, however, at about 70-80 K it experiences an abrupt jump from small negative to larger positive values. This jump coincides with the appearance of a paramagnetic phase near the interface with the substrate (see Figs. 4b and 5a). Note that the spin pinning at the ferromagnet/paramagnet interface is stronger than that either at the ferromagnet/substrate interface or at a free surface.

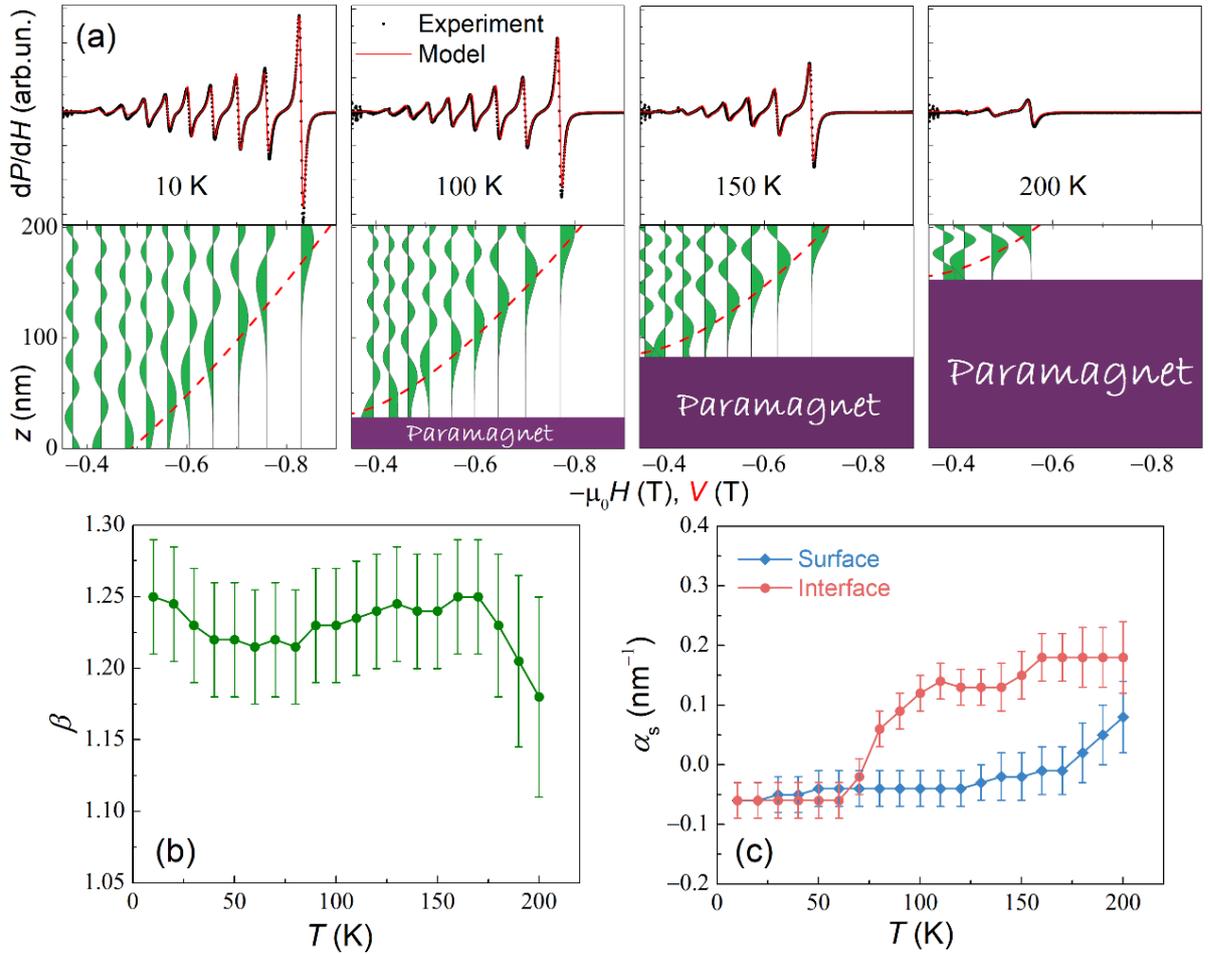

**Figure 5.** SSWR spectra of the graded Pd-Fe alloy film Lin_2-10 measured at different temperatures – top panels in (a). Black dots are the experiment data, red lines are the modeling results. The corresponding distributions of the precession amplitude $m(z)$ across the film thickness – bottom panels in (a). Here, the red dashed line is the dependence of the "potential" $V(z,T)$, given by Eq. (2), on the coordinate $z$ at the temperatures indicated above. Temperature dependences of the ratio of the effective magnetization to the saturation magnetization $\beta$ (b) and the pinning coefficients at the surface and the ferromagnet/paramagnet interface (c).

*Sample Lin_12-18*

Figure 6a shows the experimental temperature evolution of the SSWR spectrum of the Lin_12-18 graded Pd-Fe film. The spectrum temperature evolution looks more complex compared to that of the Lin_2-10 sample. Below 80 K, all sufficiently intense resonance lines (four to five modes) are observed in a narrow magnetic field range of 0.85-0.97 T. This is due to a small relative change in magnetization across the thickness, which manifests itself in a "shallow potential well" $V(z,T)$ covering a restricted magnetic field range (Fig. 6c, red dashed line in the lower panels). Absorbing spin-wave modes are located exactly within this range of fields. Modes at lower magnetic fields fill a rectangular potential well with sharp boundaries at both surfaces of the film. Such short wavelength modes weakly coupled to a microwave field show low absorption.

As the temperature increases, the well $V(z,T)$ becomes deeper (Fig. 6c, red dashed lines), and the emerging shorter-wavelength modes exhibit stronger absorption of the microwave field due better coupling. The maximum number of SSWRs is achieved at a temperature of 260 K and amounts to 11 modes (see Fig. 6a). With a further increase in temperature, the number of modes decreases, which is due to a decrease in the actual thickness and magnetization of the ferromagnetic layer, similar to the Lin_2-10 sample. It is worth noting that with increasing temperature, a surface mode (SM) descends from high magnetic fields. It emerges because of a strong pinning at the film/substrate interface [38], and its temperature dependence is determined by a change in the magnetization gradient, *i.e.*, its increase with increasing temperature. When the film region near the substrate becomes paramagnetic, the pinning strength changes (Fig. 6e), and the surface mode disappears (Figs. 6a (experiment) and 6b (modelling)).

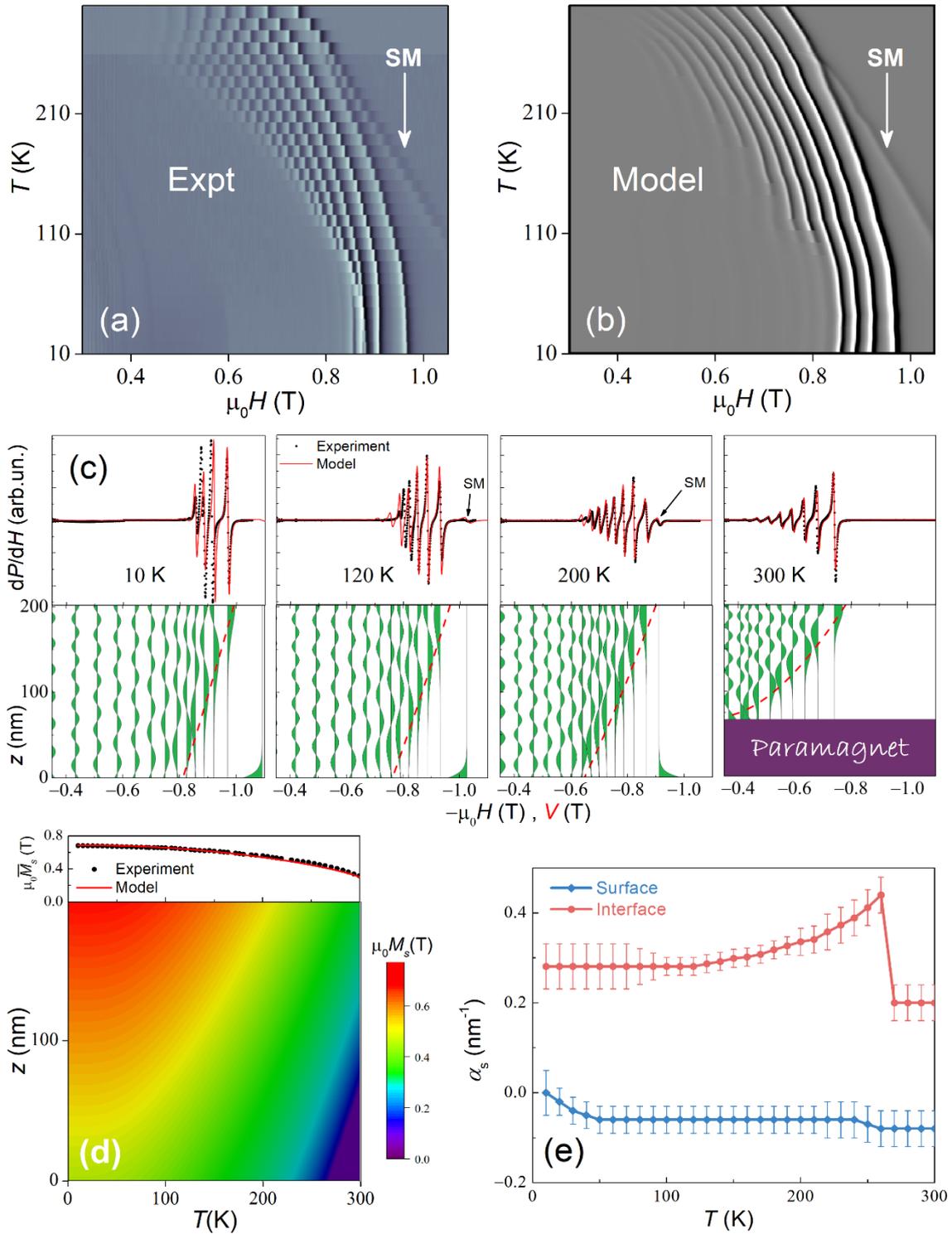

**Figure 6.** Experimental (a) and calculated (b) temperature dependences of the SSWR spectra of the Lin_12-18 sample. SSWR spectra of the graded Pd-Fe alloy film at different temperatures and the corresponding distributions of the precession amplitude *m(z)* across the film thickness (c). Distribution of the saturation magnetization across the film thickness and its variation with temperature (d), and pinning coefficients at the surface and interface of the film (e) at various temperatures. Surface mode is depicted as SM.

*Sample Sin_2-10*

The Sin_2-10 sample in terms of Eqs. (1) and (2) reveals two very different potential wells, a "deep" and a "shallow" one (Fig. 7d). The full temperature dependence of the SSWR spectrum (Fig. 7a) helps us to associate each exact mode with one of the two wells. The "deep" potential well is filled by symmetric modes with high absorption intensity and antisymmetric modes with low absorption intensity (shown by green and gray amplitude distributions, respectively, in the bottom panels of Fig. 7c). Only one mode resides in the "shallow" potential well (marked as the lone mode, LM), which almost coincides in its resonance field with the $5^{th}$ mode of the "deep" well though significantly exceeds its intensity (Figs 7b and 7c). In a finite temperature range above 75 K, the "shallow" potential well (when exists) is separated from the "deep" one by a paramagnetic layer (the violet color area in Figs. 7c and 7d). Above 150 K, the "shallow" well region loses ferromagnetism, and we observe the temperature evolution of filling in the practically symmetric "deep" well with spin waves. Pinning at both ferromagnetic region boundaries is weak (Fig. 7e).

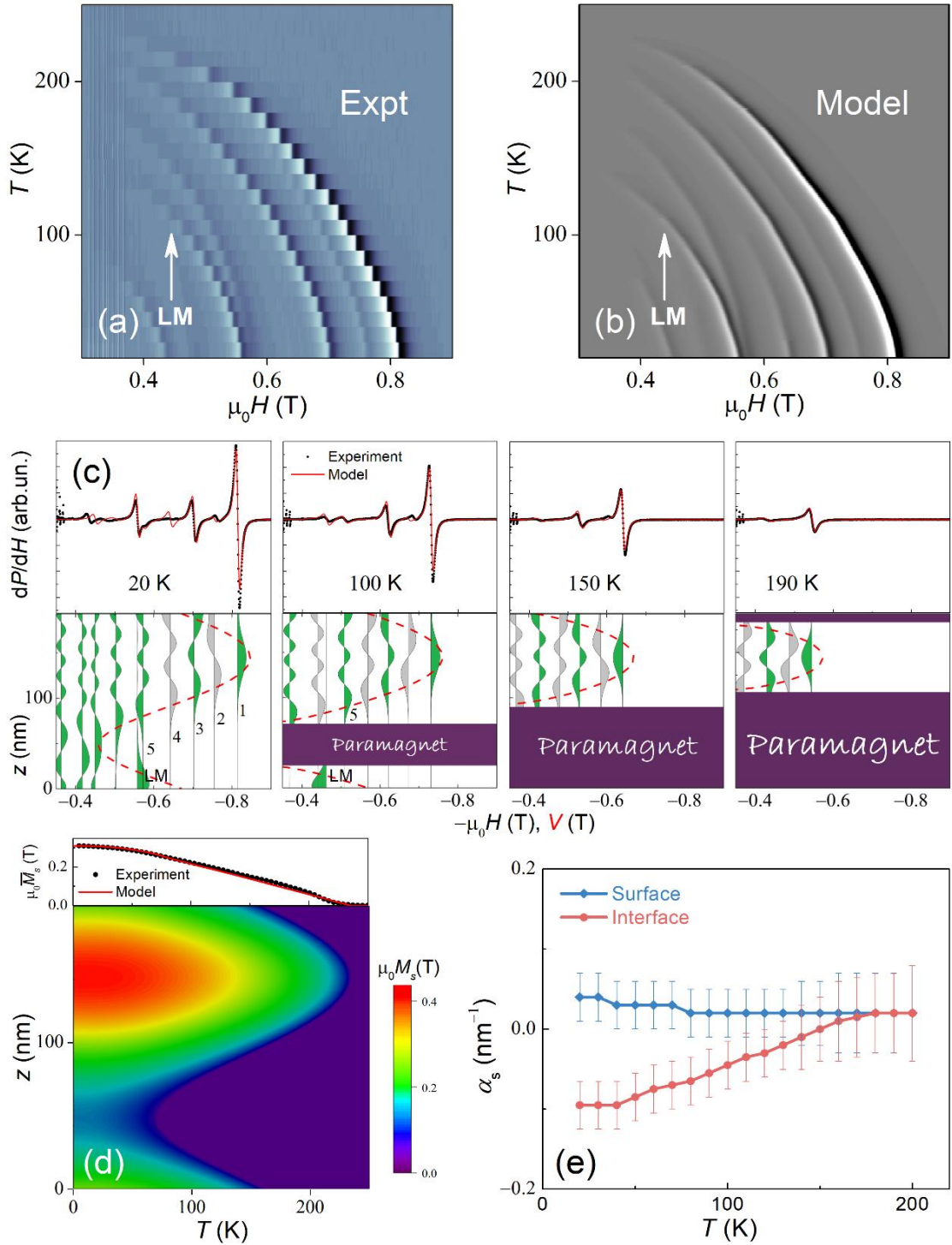

**Figure 7.** Experimental (a) and calculated (b) temperature dependences of SSWR spectra of the Sin_2-10 sample. SSWR spectra of the Sin_2-10 graded Pd-Fe film at different temperatures and the corresponding distributions of the precession amplitude $m(z)$ across the film thickness (c). Distribution of the saturation magnetization across the film thickness (d), and pinning coefficients at the surface and ferromagnet/paramagnet interface (e) at various temperatures.

*Sample Cos_2-10*

The Cos_2-10 sample at every temperature possesses two nearly identical potential wells, which is reflected in observation of resonance doublets in SSWR spectra (Figs. 8a and 8c). Based on the best fitting of experimental data on SSWR spectra, a difference in two wells is due to a slight asymmetry of the magnetization profiles relative to the central plane of the film (Fig. 8d) and difference in pinning coefficient values at the surface and the interface with the substrate (Fig. 8e). The interesting property of the temperature evolution of the SSWR spectrum of the Cos_2-10 sample is the appearance of an inflection in the temperature dependence of the resonance fields of the first four modes (Figs. 8a and 8b). This fact is associated with the splitting of a continuous ferromagnetic film into two weakly coupled magnetic layers, which manifests itself in a decrease in $\beta$ (see, Fig. 9a below, blue lines and symbols).

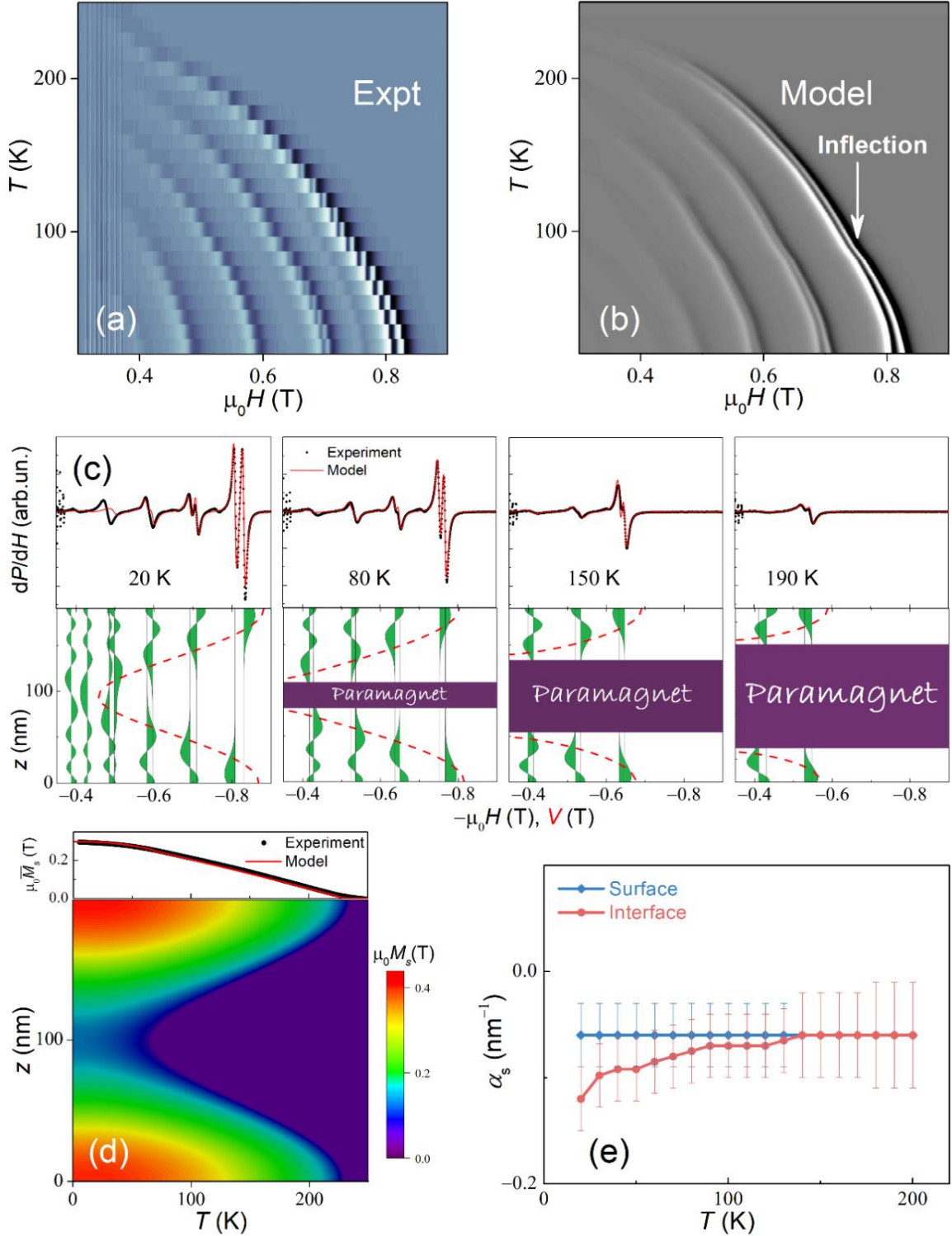

**Figure 8.** Experimental (a) and calculated (b) temperature dependences of the SSWR spectra of the Cos_2-10 sample. SSWR of the graded Cos_2-10 Pd-Fe film at different temperatures and the corresponding distributions of the precession amplitude $m(z)$ across the film thickness (c). Distribution of the saturation magnetization across the film thickness (d) and pinning coefficients at the surface and the ferromagnet/paramagnet interface (e) at various temperatures.

**Discussion**

Switching to a discussion, we state first that an expected possibility of an SSWR spectrum adjustment with temperature indeed clearly manifests itself in the presented experiments. The primary role in the observed temperature evolution of SSWR spectra is played by the temperature dependence of the magnetization represented, however, not simply by a variation of the magnetization magnitude in a certain temperature range, but more significantly by vanishing of the magnetization in a fraction of the inhomogeneous film thickness. In a studied series of samples, such transition leads to a reduction of a ferromagnetic layer thickness in one case, and to splitting of a magnetic film into two layers separated by a paramagnetic (PM) spacer in the other. The latter results in a drastic change of a size and a shape of the potential well which comprises spin-wave modes and determines the number and energies (frequencies, resonance fields) of SSWRs.

It is extremely important that there is a model proven to describe adequately standing exchange spin-waves in our samples in all implemented conditions. Then, we have at our disposal a predictive and flexile instrument which works pretty well and visualizes spin-wave modes and their frequencies just like tones of a zither/kantele of the flexible shape and size. With such an instrument, one may realize a particular spectrum at a given operation temperature starting from the model and implementing it in the material.

For our representative Lin_2_10 sample, the observed SSWR spectrum evolution is described by the modeling remarkably well. It is noteworthy here that the partial, even in a significant fraction of the thickness, transition of the material from the ferromagnetic (FM) to the PM state leads to a radical rearrangement of the spin-wave spectrum: a drop of the number of the excited modes and their resonance fields (Fig. 5a).

For the Sin_2-10 sample, spin waves with notably different properties from two strongly different "potential wells" manifest themselves in the collected spectra. At the elevated temperatures, the "shallow well" of the two vanishes, and the rather complex spectrum of excitations from the double-well potential gradually degenerates to a single mode from the residue of the former "deep well" (Fig. 7c).

In the Cos_2-10 sample, even despite a precise control of the graded film synthesis process and nominally symmetric two-well potential, both at low temperatures (when the film is completely FM) and at elevated ones (when the iron-concentrated near-interface regions become two FM layers separated by a PM spacer), pairs of spin-wave resonances with slightly different resonance fields appear (Fig. 8c). We associate these differences in frequencies with different pinning coefficients at the surface and interface of the layer.

Perhaps the experimental and modeling spectra are less affected by temperature for the Lin_12-18 film. An obvious reason for this is high concentration of iron in the alloy in the entire range of

variations – 12 to 18 at.%. The Curie temperature of the lowest concentration side of the film is close to the room temperature, therefore, below this temperature spectra depend only on the temperature variations of *M(T)* over the cross-section of the film. Moreover, the possible reason for discrepancies between the experiment and the model can be a region of high (up to 18 at.%) iron concentration, for which, due to the closeness in composition to the $Pd_3Fe$ phase [39], mesoscopic phase separation into iron-rich $Pd_3Fe$ and iron depleted regions of the solid solution could occur. This might cause deviations of the real distribution of iron across the film thickness at the high-Fe side against the technologically defined profile.

A fact that the Lin_12-18 sample to some extent falls out of a studied series follows also from the extracted temperature dependences of the model parameters *β* and *D*. Figure 9a shows the temperature dependence of the parameter *β* for all the studied samples. For samples Lin_2-10, Sin_2-10, and Cos_2-10 at low temperatures, the *β* values are all larger that unity and close to each other. With temperature increase, the evolution of the *β* parameter for the Lin_2-10 sample is slightly different from the Sin_2-10 and Cos_2-10 ones. For the sample with a linear profile, *β* does not vary significantly, while for the Sin_2-10 and Cos_2-10 samples, the *β* parameter decreases monotonously with increasing temperature. For the Lin_12-18 sample, however, the *β* parameter remains approximately equal to unity over the entire temperature range. This is consistent with our FMR measurements of a pilot homogeneous epitaxial $Pd_{0.88}Fe_{0.12}$ film (not shown here), in which the *β* parameter is also approximately equal to unity over a wide temperature range.

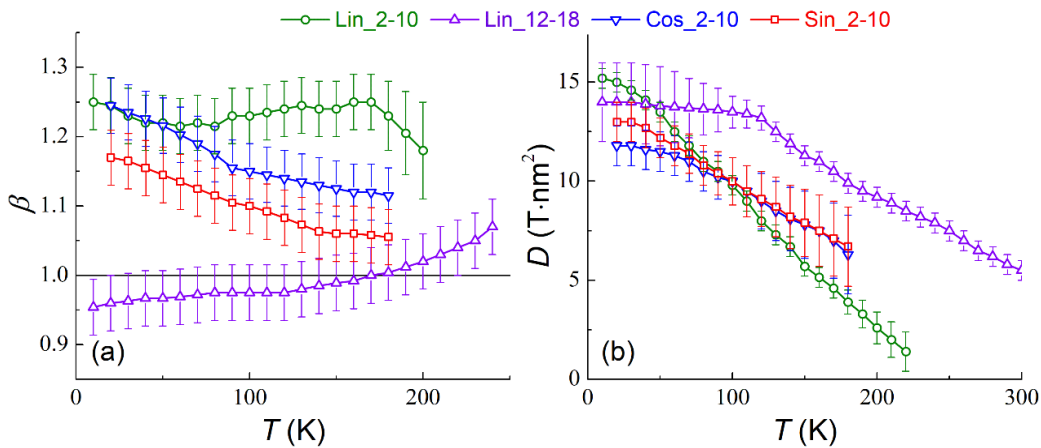

**Figure 9.** Temperature dependences of the ratio of effective magnetization to saturation magnetization *β* (a) and spin stiffness *D* (b) for samples Lin_2-10, Lin_2-18, Sin_2-10, and Cos_2-10.

The temperature dependences of the spin stiffness *D(T)* for all the studied samples are apparently determined by the dependence $D(T) \propto \bar{M}(T)$ (Fig. 9b). A similar result was obtained for a homogeneous bulk Pd-Fe single crystal using neutron studies [40]. Recall that the stiffness parameter in our modelling does not depend on iron concentration. This assumption is correct, since, based on neutron experiments on bulk Pd-Fe samples [41] and calculations [42], in the iron concentration range of 3-18 at.%, the value of *D* is only slightly concentration dependent. The spin stiffness values obtained by us are higher than those from neutron studies, where *D* at low temperatures was ≈ 9 T nm$^2$ [40,41].

Thus, the paper presents four particular and essentially different cases of graded magnetic films, the results of the experimental studies and modeling of spin-wave resonances in them. These examples, in our opinion, are sufficient to determine an impact of temperature as an easily varied experimental parameter to adjust standing spin-wave spectra of magnetically graded PdFe alloy films. It was demonstrated for the first time that control of the magnetic phase composition of the film with temperature justified by modelling leads to dramatic change in the number of the excited modes, their frequencies and shapes.

**Conclusion**

Summarizing, we have studied an evolution of standing spin wave resonance spectra in graded ferromagnetic films with temperature, which is considered as a tool for a flexile modification of the graduation profile. Four vertically graded epitaxial samples with deep modulation of the Pd-Fe alloy composition were synthesized, in particular, two with linear (2-10 at.% and 12-18 at.% range of the iron content), one with the sine and one with the cosine (both with the 2-10 at.% range of the iron content) distribution profiles across the films thickness of about 200 nm. Versatile SSWR patterns with drastic temperature variation were observed and described with high accuracy utilizing classic Landau-Lifshitz-Gilbert approach to the spin dynamics for all samples in the entire temperature range of 20-300 K. The successful numerical modeling of SSWR spectra allowed to conclude that the change in the magnetic phase composition with temperature is a main cause of the dramatic variation in the number of the excited modes, their frequencies and shapes. Consequently, a combination of the SSWR modeling and knowledge of material properties may be considered and implemented into a design of graded magnetic films with a desired SSWR spectrum at an operation temperature. Magnetically graded Pd-Fe alloy system with low (less than ~ 12-14 at.%) iron concentration is a material of a choice for such engineering.

**Acknowledgments**

L.R.T. acknowledges the support from the MegaGrant of the Ministry of Science and Higher Education of Russian Federation (No. 075-15-2024-632). The work of I.V.Y., A.I.G., B.F.G., and R.V.Y. was supported by the Kazan Federal University Strategic Academic Leadership Program (PRIORITY-2030). The authors acknowledge continuous support from I.A. Golovchanskiy during the course of the work.